# Magnetooptical determination of a topological index

Badih A. Assaf[1]*, Thanyanan Phuphachong[2]*, Valentine V. Volobuev[3,4], Günther Bauer[3], Gunther Springholz[3], Louis-Anne de Vaulchier[2], Yves Guldner[2]

[1] *Département de Physique, Ecole Normale Supérieure, CNRS, PSL Research University, 24 rue Lhomond, 75005 Paris, France*

[2] *Laboratoire Pierre Aigrain, Ecole Normale Supérieure, CNRS, PSL Research University, Université Pierre et Marie Curie, Université Denis Diderot, 24 rue Lhomond, 75005 Paris, France*

[3] *Institut für Halbleiter und Festkörperphysik, Johannes Kepler Universität, Altenberger Straβe 69, 4040 Linz, Austria*

[4] *National Technical University "Kharkiv Polytechnic Institute", Frunze Str. 21, 61002 Kharkiv, Ukraine*
\* Authors contributed equally to this work

**Abstract: When a Dirac fermion system acquires an energy-gap, it is said to have either trivial (positive energy-gap) or non-trivial (negative energy-gap) topology, depending on the parity ordering of its conduction and valence bands. The non-trivial regime is identified by the presence of topological surface or edge-states dispersing in the energy gap of the bulk and is attributed a non-zero topological index. In this work, we show that such topological indices can be determined experimentally via an accurate measurement of the effective velocity of bulk massive Dirac fermions. We demonstrate this analytically starting from the Bernevig-Hughes-Zhang Hamiltonian (BHZ) to show how the topological index depends on this velocity. We then experimentally extract the topological index in $Pb_{1-x}Sn_xSe$ and $Pb_{1-x}Sn_xTe$ using infrared magnetooptical Landau level spectroscopy. This approach is argued to be universal to all material classes that can be described by a BHZ-like model and that host a topological phase transition.**

The concept of band topology in condensed matter systems has revolutionized our understanding of quantum phases of matter.[1,2,3,4] Fundamentally speaking, it has allowed us to understand how unconventional novel states of matter can emerge in systems where fundamental symmetries are preserved. Typically, the topological character of solids is governed by the orbital and parity ordering of the conduction and valence bands. In a number of materials, band topology is proven to be non-trivial, as the parity of the conduction and valence bands is inverted compared to conventional semiconductors and the sign of the energy gap is said to be negative (Fig. 1(a)). [2,4,5,6,7] Such energy states of matter are typically attributed a non-zero topological index, that is related to the sign of the energy gap. Under certain symmetry considerations (time reversal symmetry, crystalline symmetries, etc.), systems that have a non-zero topological index host gapless metallic Dirac surface states that disperse in the band gap of the semiconductor.[2,5,6,7,8] These topological surface states (TSS) exhibit striking properties such as a relativistic energy-momentum dispersion, a helical spin texture with electron spin locked to the momentum and a high robustness against disorder.[3] The TSS give rise to novel physical effects such as quantized conductance without magnetic fields,[8] the quantum anomalous Hall effect [9,10] and may provide a basis for the realization of Majorana fermions.[11,12] A considerable number of topological material classes have been thus far identified. Among all, in the $Z_2$ topological insulator (TI) class, the band inversion occurs at an odd number of time-reversal symmetric points [13,14] in the first Brillouin zone (BZ) whereas in topological crystalline insulators (TCI) it occurs at an even number of crystalline mirror symmetric points.[15,16,17,18,19]

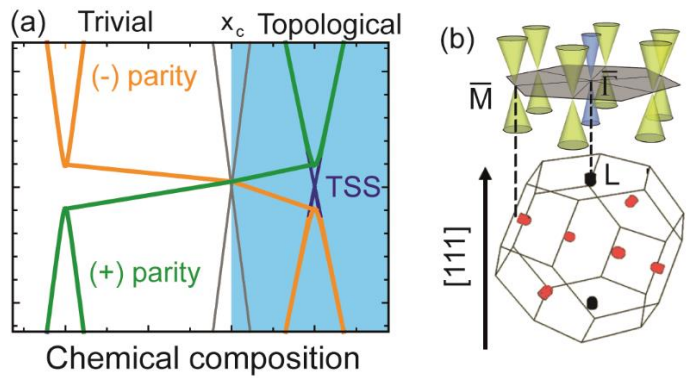

**Figure 1**. **The topological phase transition and Brillouin zone of $Pb_{1-x}Sn_xSe$ and $Pb_{1-x}Sn_xTe$.** (a) Sketch of a topological phase transition that occurs at a critical point $x_c$ as a function of composition in a system having a conduction and valence band of opposite parity (different color). The bulk band gap closes at $x_c$. Topological surface states (TSS) emerge in the topological regime $x>x_c$ (shaded blue region). In IV-VI TCI, this occurs at a critical Sn concentration $x_c$. (b) Bulk Brillouin zone of [111] IV-VI semiconductors, showing the longitudinal (black), and oblique (red) valleys. The topological surface band structure above the bulk BZ shows the $\overline{\Gamma}$ (blue) and the $\overline{M}$ (yellow) Dirac cones.

Both the TI[8,20,21,22,23] and TCI[17,18,24,25,26,27,28,29] states have been thoroughly studied experimentally. Moreover, the topological phase transition (Fig. 1(a)) from trivial to non-trivial[5,30,31,32,33,34,35,36] has also been extensively investigated for both TIs and TCI. However, no direct measurement of a topological index in 3D condensed matter systems has yet been reported. Typically, the topological index is inferred from angle resolved photoemission spectroscopy measurements that observe helical Dirac surface states in 3D TI, or via the observation of the quantum spin Hall effect in 2D TI. In materials where a topological phase transition from trivial to non-trivial can be induced, the topological index can be also inferred from the observation of a closure and reopening of the energy gap. A direct measurement of the topological index would, however, be interesting to consider fundamentally, as it may allow us to shed light on pending issues concerning the thermodynamics of topological phases. For instance, are topological phase transitions first or second order in nature?[30,] Does there exist a gapless 3D Dirac state at the critical inversion point? [30,31,37] What mechanism causes the band-inversion and the gapping of surface-states when the system becomes trivial? Can the gapping of surface-states and their acquisition of mass be described by an analogue of the Higgs mechanism? [31,36]

**Preview**

In this work, we show that the topological index can be directly measured in systems exhibiting topological phase transitions using infrared (IR) Landau level spectroscopy. Starting from the Bernevig-Hughes-Zhang (BHZ) Hamiltonian, we show that to order k² an energy dispersion that is identical to that of massive Dirac fermions can be obtained, with a modified Dirac velocity that depends on the topological index. Using magnetooptical IR Landau level spectroscopy, we are able to measure this effective velocity with high precision in TCI $Pb_{1-x}Sn_xSe$ and $Pb_{1-x}Sn_xTe$ and thus experimentally extract the topological index. This is a first proof of concept that topological indices can be experimentally measured, and not just inferred from the observation of edge/surface states. We argue that our result is expected to be valid for other systems that exhibit a topological phase transition such as (Hg,Cd)Te,[38] $Cd_3As_2$,[39] or $BiTlS_{1-\delta}Se_\delta$.[36,35]

**Theory**

**The topological index in the BHZ Hamiltonian.** Let us start by solving the eigenvalue problem for a system described by the general[40] BHZ effective Hamiltonian.[2] Since we define z to be the direction of the applied magnetic field, cyclotron motion in the plane perpendicular to the field will be considered. We can thus set $k_z=0$, which gives:

$$H(k, k_z = 0) = \begin{pmatrix} \Delta - M_1 k^2 & 0 & 0 & \hbar v_c k_- \\ 0 & -\Delta + M_1 k^2 & \hbar v_c k_- & 0 \\ 0 & \hbar v_c k_+ & \Delta - M_1 k^2 & 0 \\ \hbar v_c k_+ & 0 & 0 & -\Delta + M_1 k^2 \end{pmatrix} \quad (1)$$

Here, $\Delta = E_g/2$ is the half-band gap, $v_c$ is a velocity usually related to the **k.p** matrix element and $M_1$ ($M_1<0$ in the sign convention of BHZ)[40] is a Schrödinger mass term. $k_\pm = k_x \pm i k_y$, $k^2 = k_x^2 + k_y^2$. The exact eigen-energies for the conduction and valence bands $E_{C/V}$ resulting from this Hamiltonian are given by:

$$E_{C/V} = \pm\sqrt{(\Delta - M_1 k^2)^2 + (\hbar v_c k)^2} \quad (2)$$

Neglecting $k^4$ terms, this can be simplified to:

$$E_{C/V} = \pm\sqrt{\Delta^2 + \left(v_c^2 - \frac{2M_1\Delta}{\hbar^2}\right)\hbar^2 k^2} \quad (3)$$

Or,

$$E_{C/V} = \pm\sqrt{\Delta^2 + v_D^2 \hbar^2 k^2} \quad (4)$$

This is essentially a massive Dirac dispersion having an effective Dirac velocity $v_D^2 = v_c^2 - \frac{2M_1\Delta}{\hbar^2}$. In this notation $v_c$ is the critical velocity of a gapless 3D Dirac state that may exist at the critical point between the trivial and non-trivial phases. Interestingly the term,

$$v_D^2 - v_c^2 = -\frac{2M_1\Delta}{\hbar^2} \qquad (5)$$

is related to the topological index $\eta$ as previously defined in the literature $(-1)^\eta = sign(-\frac{2M_1\Delta}{\hbar^2})$.[34,41,42]

Measuring $v_D$ and $v_c$ can thus yield a measure of the topological index $\eta$ (modulo 2).

Note that, even if one cannot neglect $k^4$, Eq. (4) can still be expressed in terms of a Dirac velocity that yields the topological index. This is discussed later on in the text.

**The case of Pb$_{1-x}$Sn$_x$Se and Pb$_{1-x}$Sn$_x$Te.** In order to experimentally study the topological phase transition, and attempt to measure the topological index, we investigate the case of Pb$_{1-x}$Sn$_x$Se and Pb$_{1-x}$Sn$_x$Te IV-VI TCI, where the transition occurs (Fig. 1(a)) as a function of changing Sn content, at four L-points in the Brillouin zone (Fig. 1(b)).[30,31,43,44,45] The bulk Fermi surface of Pb$_{1-x}$Sn$_x$Se and Pb$_{1-x}$Sn$_x$Te is shown in Fig. 1(b). It is degenerate and possesses an ellipsoidal bulk carrier valley oriented parallel to the [111] axis referred to as the longitudinal valley as well as the three tilted valleys referred to as the oblique valleys. The four-fold bulk valley degeneracy combined with the mirror symmetric character of the rock-salt crystal structure of Pb$_{1-x}$Sn$_x$Se and Pb$_{1-x}$Sn$_x$Te yields a TCI state that hosts four-fold degenerate TSS.[5,16,25,46,47,48,49,50,51]

Despite this subtlety, the band structure and Landau levels of IV-VI semiconductors is ideal to study under the proposed scope, due to their relative simplicity[24,52] compared to other materials such as II-VI and V-VI systems that have highly asymmetric conduction and valence bands. It can actually be shown that a description similar to that of BHZ can be applied to describe the band structure of Pb$_{1-x}$Sn$_x$Se and Pb$_{1-x}$Sn$_x$Te. Starting from a 2-band **k.p** Hamiltonian where $\Delta$ is defined to be the half-band-gap again, and $v_c = \frac{P_{k.p}}{m_0}$ where $P_{k.p}$ is the **k.p** matrix element of each respective valley,[45] we can then perturbatively introduce contributions from far-bands as $M_1 = -\frac{\hbar^2}{2\tilde{m}}$ where $\tilde{m}$ is a far-band correction mass term.[43,45,52,53,54,55,56] Interestingly, in Pb$_{1-x}$Sn$_x$Se and Pb$_{1-x}$Sn$_x$Te, $M_1$ turns out to be relatively small.[57] This theoretical treatment then yields a massive Dirac-like energy dispersion identical to (4) with an effective Dirac velocity given by:

$$v_D^2 = v_c^2 + \frac{\Delta}{\tilde{m}} \qquad (6)$$

This is equivalent to writing the band-edge mass as:

$$\frac{1}{m} = \frac{v_c^2}{|\Delta|} + \frac{1}{\tilde{m}}$$

Δ changes sign through the topological phase transition, as the conduction and valence bands swap, however, $\widetilde{m}$ does not change sign. Generally, $\widetilde{m}$ is due to interactions between the valence/conduction band states, and other bands lying far away from the band gap in energy, referred to as far-bands. The ordering of far-bands does not invert when the valence and conduction bands invert, thus the sign of $\widetilde{m}$ does not change for fundamental reasons.[45, 58]

The Landau level spectrum in this case can be determined via the standard procedure of performing a Peierls substitution and using a ladder operator formalism, following what is done is ref. 54. This is detailed in supplementary section 1; we get:

$$E_{N>0}^{c,\pm} = \mp\hbar\widetilde{\omega} + \sqrt{\Delta^2 + 2v_D^2\hbar eBN} \quad (7a)$$
$$E_0^c = \hbar\widetilde{\omega} + \Delta$$

$$E_{N>0}^{v,\pm} = \pm\hbar\widetilde{\omega} - \sqrt{\Delta^2 + 2v_D^2\hbar eBN} \quad (7b)$$
$$E_0^v = -\hbar\widetilde{\omega} - \Delta$$

Here, N is the Landau level index and $\widetilde{\omega} = eB/\widetilde{m}$. ± refers to the effective spin. We highlight that this result is analytically equivalent to the result obtained by Mitchell and Wallis in 1966,[54] and subsequent work on the Landau levels of PbTe and other lead-tin-salts.[45,57] Mitchell and Wallis also neglected terms in $B^2$ which essentially yields the same result that we have after neglecting $k^4$ contributions.

For $Pb_{1-x}Sn_xSe$ and $Pb_{1-x}Sn_xTe$, we can then evaluate the topological index, as defined by Fu[6] and later used by Juricic et al.[34] for TCIs, as:

$$(-1)^\eta = sign(\frac{\Delta}{\widetilde{m}}) \quad (8)$$

$$\text{Where, } \frac{\Delta}{\widetilde{m}} = v_D^2 - v_c^2 \quad (9)$$

$\eta = 1$ corresponds to a non-trivial phase and $\eta = 0$ to the trivial phase. Here $\eta$ is a valley topological index that can be related to the mirror Chern number in TCI via the definition given by Hsieh[5] et al. and Fu.[6,59]

**Experimental Results**

**Growth and characterization.** In order to study the topological transition in $Pb_{1-x}Sn_xSe$ and $Pb_{1-x}Sn_xTe$, epilayers are grown in the [111]-direction by molecular beam epitaxy (MBE) on freshly cleaved $BaF_2$ (111) substrates.[60,61,62] The Sn concentration x is systematically varied over a wide range, 0≤x≤0.3 for $Pb_{1-x}Sn_xSe$ and 0≤x≤0.56 for $Pb_{1-x}Sn_xTe$. *In-situ* reflection high energy electron diffraction and *ex-situ* atomic force microscopy are initially used to characterize the films (see supplementary material S2). X-ray diffraction (XRD) (Fig. 2) is used in order to determine the lattice constant and composition with a precision better than 2%. Fig. 2(a,c) shows the (222) XRD Bragg reflection for a series of $Pb_{1-x}Sn_xSe$ (Fig. (2a)) and $Pb_{1-x}Sn_xTe$ (Fig. (2c)) films with different Sn content, illustrating the monotonic shift of the (222) diffraction peaks to higher diffraction angles with increasing Sn content. Epilayers having a thickness > 0.5 µm are proven to be fully relaxed by studying reciprocal space maps showing the (513) Bragg reflection.[24] From the peak positions, the lattice constant $a_0$ and Sn concentration of the ternary material can be directly obtained from Vegard's law (Fig. 2(b)).

Transport measurements are performed to extract the carrier density and mobility. Carrier densities as low as $10^{17} cm^{-3}$ are achieved in $Pb_{1-x}Sn_xSe$. For $Pb_{1-x}Sn_xTe$ x>0.25, moderate Bi doping (<$10^{19}cm^{-3}$)[63] is used to limit the carrier density to no more than p=$2x10^{18}cm^{-3}$. The Hall mobility μ of the samples at 77K is measured to be around 30000 to 60000 $cm^2$/Vs for $Pb_{1-x}Sn_xSe$, and between 5000 and 20000 $cm^2$/Vs for $Pb_{1-x}Sn_xTe$. These excellent transport properties allow us to observe Landau quantization at low magnetic fields.

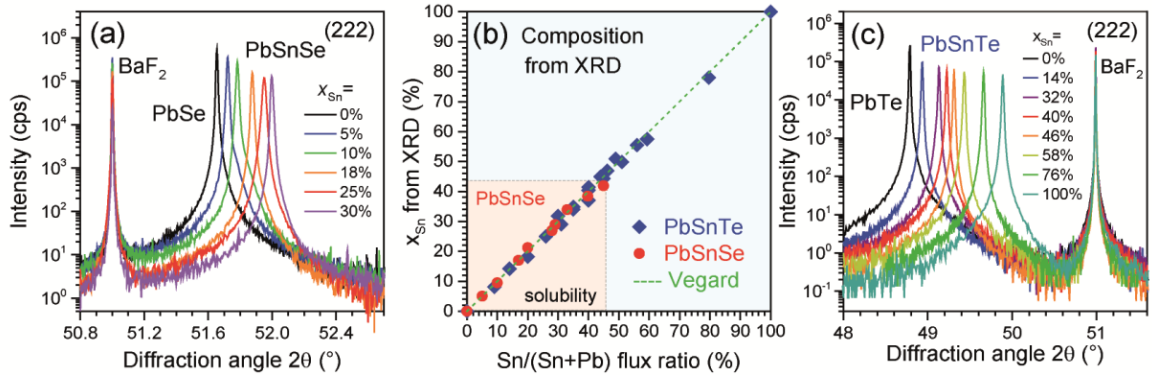

**Figure 2. X-ray diffraction data and analysis.** XRD characterization of (a) $Pb_{1-x}Sn_xSe$ and (c) $Pb_{1-x}Sn_xTe$ epilayers on $BaF_2$ (111) substrates with Sn content $x_{Sn}$ varying from 0 to 0.3 and 0 to 1 respectively. From the change in lattice constant, the Sn concentration of the layers was derived using the Vegard's law. The result is shown in (b) where the composition determined by X-ray diffraction is plotted versus the measured beam flux ratio Sn/(Sn+Pb) used for growth in the case of $Pb_{1-x}Sn_xSe$ (●) and $Pb_{1-x}Sn_xTe$ (◆). As indicated by the dashed line, the data points agree very well with the nominal values. The boundaries of the Sn solubility limit in single phase cubic $Pb_{1-x}Sn_xSe$ of about 0.4 is indicated.

**Magnetooptical Landau-level spectroscopy of the bulk band structure.** Magnetooptical IR Landau level spectroscopy is then performed in transmission mode in the Faraday geometry up to 17T, at T=4.5K in the mid-IR range. The relative transmission at fixed magnetic field T(B)/T(B=0) is extracted and analyzed. This technique is highly sensitive to the bulk, yet not blind to the TSS, and is thus an ideal tool to study the inversion of bulk energy bands in topological materials to measure the topological index. Additionally, this technique allows a quantitative assessment of the TSS band structure, via cyclotron resonance measurements.[24]

Landau level transitions Fig. 3(a,b) and magnetooptical spectra Fig. 3(c,d) are shown for two $Pb_{1-x}Sn_xSe$ samples that are respectively trivial (x=0.14) and non-trivial (x=0.19) at T=4.5K.[44,30] Minima observed in Fig. 3(c,d) correspond to absorptions due to the presence of Landau-level (LL) transitions at high magnetic fields (μB>>1). Transitions from a valence band LL to a conduction band LL are referred to as interband transitions. Transitions between two LL of the same band are referred to as intraband transitions, or more simply as cyclotron resonances (CR). A large number of interband transitions is observed (Fig. 3(a,b)) down to low energies evidencing a Fermi level position close to the valence band edge in all samples. Sharp absorption lines can be seen in all spectra with a field onset close to 1T, evidencing a very high mobility. An energy cutoff at 55meV and down to 22meV in the far-IR is due to the Reststrahlen band of the $BaF_2$ substrate.

Due to the many-valley band structure of [111]-oriented IV-VI semiconductors, two Landau level series, (Fig. 3(a,b)) pertaining to the longitudinal (black) and oblique (red) valleys (see Fig. 1(b)) are identified and analyzed for B||[111]. The oblique valleys are tilted by 70.5° with respect to the longitudinal one.[64] The oblique velocity is lower than the longitudinal velocity. This anisotropy in $v_D$ is small for $Pb_{1-x}Sn_xSe$ but rather large for $Pb_{1-x}Sn_xTe$ (Supplement (S3)).[24] The interband transitions in Fig. 3(a,b) can be well described using a massive Dirac-like Landau-level spectrum given in Eq. (7). For $k_z=0$ ($k_z // B$), the selection rules for the Faraday geometry result in the transition energies given by:[54]

$$E_N^c - E_{N\pm 1}^v = \sqrt{\Delta^2 + 2e\hbar N v_D^2 B} + \sqrt{\Delta^2 + 2e\hbar (N\pm 1)v_D^2 B} \quad (10a)$$

The ground cyclotron resonance energy at $k_z=0$ is given by:[53]

$$E_{CR} = \sqrt{\Delta^2 + 2e\hbar v_D^2 B} - |\Delta| \quad (10b)$$

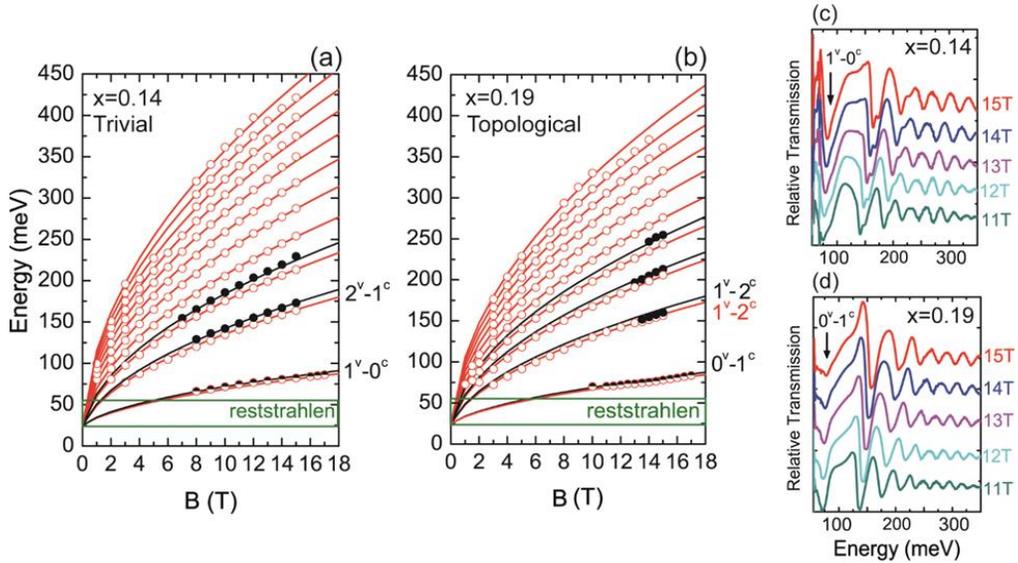

**Figure 3**. **Magnetooptical transition fan-charts and transmission spectra for $Pb_{1-x}Sn_xSe$.** (a,b) Magnetooptical Landau-level transitions for $Pb_{1-x}Sn_xSe$ x=0.14, and x=0.19 at T=4.5K. Interband transitions from the valence to the conduction band are labeled [$N^v$–$(N\pm 1)^c$]. Circles denote data points extracted from mid-IR spectra partially shown in (c,d). Solid lines are curve fits using the massive Dirac model for bulk states Eq. (10a). Black and red are used for the longitudinal and oblique bulk valleys respectively. The green frame is the Reststrahlen band of $BaF_2$. (c,d) Mid-IR magnetooptical transmission spectra measured at different magnetic fields for x=0.14 and x=0.19. Curves are shifted for clarity.

The curve fits in Fig. 3(a,b) allow us to precisely determine the energy gap $|E_g|=|2\Delta|$ and $v_D$ for all compositions. For trivial $Pb_{0.86}Sn_{0.14}Se$, we find $|2\Delta|$=25±5meV, and $v_D$=(5.05±0.10)x10$^5$m/s for the longitudinal and $v_D$=(4.8±0.1)x10$^5$m/s for the oblique valleys, whereas for non-trivial $Pb_{0.81}Sn_{0.19}Se$ we find $|2\Delta|$=25±10meV and $v_D$=(4.8±0.1)x10$^5$m/s for the longitudinal and $v_D$=(4.6±0.1)x10$^5$m/s for the oblique valleys. The Fermi level can be extracted from the onset of the $1^v$-$0^c$ transition and the CR. We find 20meV and 25meV from the band edge in x=0.14 and x=0.19 respectively, in agreement with the bulk carrier density (p≈2x10$^{17}$cm$^{-3}$ and n≈3x10$^{17}$cm$^{-3}$, respectively).

**Measurement of the topological index.** We systematically study a total of 8 $Pb_{1-x}Sn_xSe$ samples (0≤x≤0.3) and 20 $Pb_{1-x}Sn_x$Te samples (0≤x≤0.56). Note that in all samples, the magnetooptical transitions can be well interpreted with a massive Dirac model (Eq. 7) in order to extract $|2\Delta|$ and $v_D$. The analysis is shown in Fig. 4 for $Pb_{1-x}Sn_xSe$ and in the supplement (S3) for $Pb_{1-x}Sn_x$Te. The longitudinal and oblique velocities extracted from the massive Dirac model are respectively plotted versus x in Fig. 4(a,b). They show a consistent decrease as x is increased for both materials. The band gap $|2\Delta|$ is shown in Fig. 4(c). A minimum is observed for x=0.165 followed by an increase when x is increased beyond this concentration. The smallest measured energy gap is 15±5meV.[30] We can still estimate $v_c$ the critical velocity to be equal to $5.0 \times 10^5$m/s for the longitudinal valley and $4.7 \times 10^5$m/s for the oblique valleys (Fig. 4(a,b)), since the critical composition of $Pb_{1-x}Sn_xSe$ is known. In Fig. 4(d) $\Delta/\tilde{m} = v_D^2 - v_c^2$ is depicted. The sign change of $\Delta/\tilde{m}$ reflects the emergence of a topological phase in $Pb_{1-x}Sn_xSe$ having a non-zero topological index $\eta = 1$. *A first measurement of the bulk topological index in a material family of 3D topological insulators is thus confirmed in our experiment.*

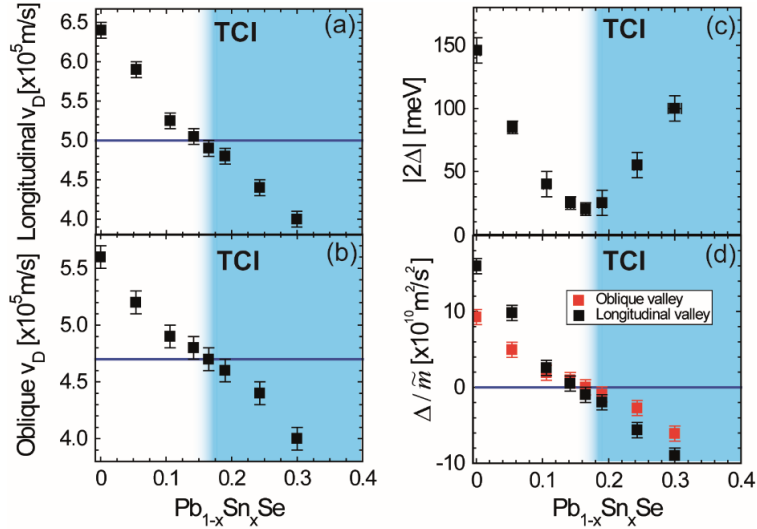

**Figure 4**. **Dependence of the velocity on the energy-gap and its sign inversion in $Pb_{1-x}Sn_xSe$.** Longitudinal (a) and oblique (b) velocity and energy-gap (c) measured in $Pb_{1-x}Sn_xSe$ at 4.5K. The blue line in (a) and (b) indicates the position of the critical velocity $v_c$. (d) $\Delta/\tilde{m}$ plotted as a function of x. Shaded blue area indicates the topological regime.

In order to verify the consistency of the approximations used in the theory, we estimate the value of $\tilde{m}$ from the measurement of the energy gap and $\Delta/\tilde{m}$ and check that its contribution is indeed small. The values are listed in table I. We discuss this more thoroughly in the supplement (S4).

| $Pb_{1-x}Sn_xSe$ | $\tilde{m}/m_0$ for oblique valleys |
|---|---|
| 0 | 0.15±0.02 |
| 0.05 | 0.16±0.02 |
| 0.10 | 0.18±0.03 |
| 0.14 | 0.23±0.03 |
| 0.19 | 0.24±0.03 |
| 0.24 | 0.18±0.03 |
| 0.30 | 0.15±0.02 |

Table I. $\tilde{m}/m_0$ for the oblique valleys determined for the investigated samples.

**Cyclotron resonance of topological surface states.** Finally, we corroborate our findings by further evidencing the observation of topological surface states that appear in the non-trivial regime. Measurements in the far-IR up to 17T are performed and are shown for x=0.14 and x=0.19 in Fig. 5(a,b). The Landau level transitions in the far-IR are shown in Fig. 5(c,d). The CR of the bulk valleys (CR-LO) and the first interband transition can be seen in $Pb_{0.86}Sn_{0.14}Se$ (Fig.5(a,c)). In $Pb_{0.81}Sn_{0.19}Se$, two transitions can be resolved, the first interband transition, as well as an additional transition marked by a blue arrow in Fig. 5(b,d). It occurs at energies higher than 60 meV where the CR-LO of the bulk bands is expected, as seen in Fig. 5(d). It is observed in $Pb_{0.81}Sn_{0.19}Se$ (Fig. 5(b)), but not in $Pb_{0.86}Sn_{0.14}Se$ (Fig. 5(a)). Its dispersion (blue in Fig. 5(d)) agrees well with that of Landau-levels of massless Dirac fermions having a velocity $v_D$ =(4.7±0.1)x10$^5$m/s - almost equal to that of the bulk valleys:

$$E_1 - E_0 = \sqrt{2e\hbar v_D^2 B} \qquad (11)$$

This transition could thus attributed to the ground-state CR of massless Dirac TSS.

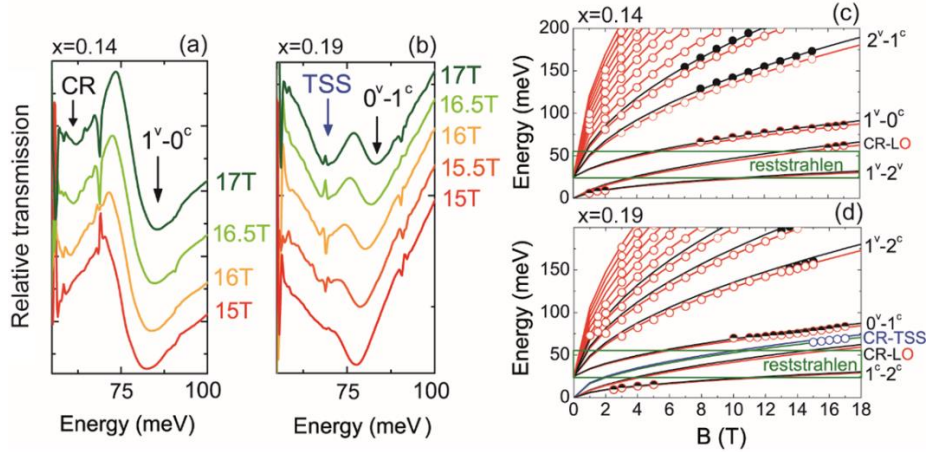

**Figure 5.** Far-IR magnetooptical spectra of samples x=0.14 (a) and x=0.19 (b). The first interband transition is labeled in both. The bulk CR is labeled in (a) and the CR-TSS is labeled in blue in (b). (c) x=0.14 and (d) x=0.19 Landau level dispersions with data points for bulk CR added in black (c) and for CR-TSS added in blue (d). Solid lines are curve fits using massive Dirac model transitions for bulk states in red and black (Eq. 10(a,b)) and massless Dirac model for CR-TSS (Eq. 11). The reststrahlen of $BaF_2$ is shown in green.

This transition is only visible at 15T and above in $Pb_{0.81}Sn_{0.19}Se$. Accordingly, the Fermi level is estimated to be around 60meV from the Dirac point. We cannot distinguish the $\overline{\Gamma}$-Dirac cone from the $\overline{M}$-Dirac cones (see Fig. 1 (b)) in $Pb_{0.81}Sn_{0.19}Se$ since the two have similar Fermi velocities and hence have overlapping CR. This might also explain the large intensity of the transition attributed to the TSS. Further evidence of the observation of massless TSS in $Pb_{1-x}Sn_xTe$ (x>0.40) is presented in the supplement (S2) as well as in a previous work on $Pb_{0.54}Sn_{0.46}Te$.[24] All in all, Fig. 5 shows that a CR resulting from TSS is observed when $\Delta/\tilde{m} < 0$.

**Discussion**

*In conclusion, our result is an analytical and experimental proof that the topological index of 3D topological insulators can be measured by magnetooptical Landau level spectroscopy.* Our approach is most suitable for ultra-narrow gap systems, having very light effective masses. It is not as suitable for materials having heavy effective masses such as $Bi_2Se_3$.

In [111]-oriented $Pb_{1-x}Sn_xSe$ and $Pb_{1-x}Sn_xTe$, we have experimentally shown that a measurement of $v_D$ relative to the critical $v_c$ allows one to experimentally determine the topological character of a material. *This work is thus a proof of concept that the topological index can be measured using a technique that quantifies the bulk band parameters, and not just inferred from the observation of surface states.* Our results are further supported by the observation of a cyclotron resonance from massless Dirac topological surface-states in the non-trivial regime. We believe this approach can be applicable to any system supporting massive or massless Dirac fermions that go through a topological phase transition and that can be described by a BHZ Hamiltonian.[36,35,38,40,65,20,66,67,68,69,70] Accordingly, we propose to further study under the same scope compound series such as $Hg_{1-x}Cd_xTe$ (see supplement 5),[38] $BiTlS_{1-\delta}Se_\delta$,[36,35] $Bi_{2-x}In_xSe_3$,[68] topological Heusler materials[69] as well as Dirac semimetals such as $Na_3Bi$ and $Cd_3As_2$,[39,70] all expected to have a trivial to non-trivial topological phase transition.

## Methods

**MBE growth.** Epitaxial growth of (111) $Pb_{1-x}Sn_xSe$ and $Pb_{1-x}Sn_xTe$[60,61] films on $BaF_2$ (111) substrates is performed using a Riber 1000 and a Varian GEN-II molecular beam epitaxy setup respectively. Samples are grown under UHV conditions better than $5 \times 10^{-10}$ mbar. Effusion cells filled with stoichiometric PbSe, PbTe, SnSe and SnTe are used as source material. The chemical composition of the ternary layers is varied over a wide range by control of the SnSe/PbSe (SnTe/PbTe) beam flux ratio that is measured precisely using a quartz microbalance moved into the substrate position. The growth rates are typically 1 μm/hour (~1 monolayer/sec) and the growth temperature is set to 380°C as checked by an IRCON infrared pyrometer. The thickness of the films is in the range of 1– 3 μm. In order to obtain a low free carrier concentration (<$10^{18}$ cm$^{-3}$), and compensate the native background hole concentration that increases strongly for higher Sn concentration in $Pb_{1-x}Sn_xSe$ and $Pb_{1-x}Sn_xTe$, extrinsic *n*-type Bi-doping was provided by $Bi_2Se_3$ or $Bi_2Te_3$ doping cells.[63] The growth is monitored in-situ using reflection high energy electron diffraction (RHEED).

**X-ray diffraction**. X-ray diffraction measurements are performed using Cu-Kα 1 radiation in a Seifert XRD3003 diffractometer, equipped with a parabolic mirror, a Ge(220) primary beam Bartels monochromator and a Meteor 1D linear pixel detector.

**Transport characterization.** Transport measurements are performed at 77K using a van der Pauw geometry in order to determine the Hall carrier density and mobility.

**Magnetooptical absorption spectroscopy.** Magnetooptical absorption experiments are performed in an Oxford Instruments 1.5K/17T cryostat at 4.5K. Spectra are acquired using a Bruker Fourier transform spectrometer. All measurements are made in Faraday geometry. Two different broadband sources are used for different IR ranges: a far-IR source (3-900cm$^{-1}$) and a mid-IR source (400-3600cm$^{-1}$). Measurements are performed at fixed magnetic fields between 0 and 15T and occasionally up to 17T. A He cooled bolometer, is used to detect the transmitted signal. The relative transmission at fixed magnetic field T(B)/T(B=0) is extracted and analyzed. Baseline signal contributions from the bolometer's response variations in the magnetic field are negligible in this experiment, due to the large amplitude of the observed transitions.

**Acknowledgements:** We would like to thank T. Kontos, V. Juricic, S.Q. Shen and X.R. Wang for useful comments. This work is supported by Agence Nationale de la Recherche LabEx grant ENS-ICFP (ANR-10-

LABX-0010/ANR-10-IDEX-0001-02 PSL) and by the Austrian Science Fund, Project SFB F2504-N17 IRON. TP acknowledges support from the Mahidol Wittayanusorn Scholarship and the Franco-Thai Scholarship.

**Author contributions**

TP, LAV and YG performed magnetooptical IR measurements. BAA, TP, LAV and YG analyzed the data with the assistance of GB. BAA and YG developed and tested the theoretical description with the assistance of GB. VVV and GS grew by MBE and characterized the samples using XRD, EDX, RHEED and transport. BAA, LAV and YG conceived the project. All authors collaboratively discussed the results and contributed to the writing of the manuscript.

The authors declare no competing financial interest.

**Supplementary Material for:**

# Magnetooptical determination of a topological index

Badih A. Assaf[1], Thanyanan Phuphachong[2], Valentine V. Volobuev[3,4], Günther Bauer[3], Gunther Springholz[3], Louis-Anne de Vaulchier[2], Yves Guldner[2]

[1] Département de Physique, Ecole Normale Supérieure, CNRS, PSL Research University, 24 rue Lhomond, 75005 Paris, France

[2] Laboratoire Pierre Aigrain, Ecole Normale Supérieure, CNRS, PSL Research University, Université Pierre et Marie Curie, Université Denis Diderot, 24 rue Lhomond, 75005 Paris, France

[3] Institut für Halbleiter und Festkörperphysik, Johannes Kepler Universität, Altenberger Strasse 69, 4040 Linz, Austria

[4] National Technical University "Kharkiv Polytechnic Institute", Frunze Str. 21, 61002 Kharkiv, Ukraine

## S1. Landau quantization in the Mitchell-Wallis description

We now will treat the problem of Landau quantization for the longitudinal valley of (111)-oriented IV-VI semiconductors using the Mitchell-Wallis (MW) description [1,2,3]. The Landau levels are given by a 6-band **k.p** approach, where the $L_6^\pm$ bands are exactly accounted for and the effect of four far-bands (two conduction and two valence bands) is included perturbatively in $k^2$-approximation. At $k_z=0$, the Landau level energies of the conduction (c) and valence (v) bands are expressed as:

$$\begin{aligned}
E_n^{c,\pm} &= \frac{1}{2}\left\{\left(n+\frac{1}{2}\right)(\hbar\widetilde{\omega}_c + \hbar\widetilde{\omega}_v) \pm \left[(\tilde{g}_c - \tilde{g}_v)\frac{\mu_B B}{2} + \hbar\widetilde{\omega}_v\right]\right\} \\
&\quad + (\Delta)^{1/2}\{[\Delta + (n+\frac{1}{2})(\hbar\omega_c - \hbar\omega_v)] \pm [(g_c + g_v)\frac{\mu_B B}{2} - \hbar\omega_v]\}^{1/2} \\
E_n^{v,\pm} &= \frac{1}{2}\left\{\left(n+\frac{1}{2}\right)(\hbar\widetilde{\omega}_c + \hbar\widetilde{\omega}_v) \mp \left[(\tilde{g}_c - \tilde{g}_v)\frac{\mu_B B}{2} - \hbar\widetilde{\omega}_c\right]\right\} \\
&\quad - (\Delta)^{1/2}\{[\Delta + (n+\frac{1}{2})(\hbar\omega_c - \hbar\omega_v)] \mp [(g_c + g_v)\frac{\mu_B B}{2} - \hbar\omega_c]\}^{1/2}
\end{aligned}$$

(S1)

Here, the cyclotron frequencies of the conduction ($\omega_c$) and valence ($\omega_v$) bands are defined respectively as $\omega_c = \omega + \widetilde{\omega}_c$ and $\omega_v = -\omega + \widetilde{\omega}_v$, where $\omega=eB/m$ is the 2-band cyclotron frequency (with $\frac{1}{m} = \frac{v_c^2}{\Delta}$). $\widetilde{\omega}_c$ and $\widetilde{\omega}_v$ are the far-band contributions. The effective g factors of the conduction ($g_c$) and valence ($g_v$) bands are given by $g_c = g + \tilde{g}_c$ and $g_v = -g + \tilde{g}_v$, where the tilde terms represent the far-band contributions. Also note that terms that vary as $B^2$ in the square root are explicitly neglected in the Mitchell and Wallis paper. This is equivalent to neglecting $k^4$ terms in the BHZ eigenvalue.

Near the topological phase transition, we can use the simplifying assumption that[4]:

$$\widetilde{\omega}_c = -\widetilde{\omega}_v = \widetilde{\omega}$$

$$\tilde{g}_c = -\tilde{g}_v = \tilde{g}$$

(S1) simplifies to

$$E_n^{c,\pm} = \pm\frac{1}{2}\{\tilde{g}\mu_B B - \hbar\tilde{\omega}\}$$
$$+(\Delta)^{1/2}\{[\Delta + (n+\frac{1}{2})2\hbar(\omega+\tilde{\omega})] \pm \hbar(\omega+\tilde{\omega})\}^{1/2}$$

(S2)

$$E_n^{v,\pm} = \mp\frac{1}{2}\{\tilde{g}\mu_B B - \hbar\tilde{\omega}\}$$
$$-(\Delta)^{1/2}\{[\Delta + (n+\frac{1}{2})2\hbar(\omega+\tilde{\omega})] \pm \hbar(\omega+\tilde{\omega})\}^{1/2}$$

Further assuming that $\tilde{g}\mu_B B \approx -\hbar\tilde{\omega}$ leads to,

$$E_n^{c,\pm} = \mp\hbar\tilde{\omega}$$
$$+(\Delta)^{1/2}\{\Delta + (2n+1\pm1)\hbar(\omega+\tilde{\omega})\}^{1/2}$$
$$E_n^{v,\pm} = \pm\hbar\tilde{\omega}$$
$$-(\Delta)^{1/2}\{\Delta + (2n+1\pm1)\hbar(\omega+\tilde{\omega})\}^{1/2}$$

Writing $2\Delta(\omega+\tilde{\omega}) = 2\Delta(eB/m + eB/\tilde{m}) = 2eB(v_c^2 + \Delta/\tilde{m})$ now gives:

$$E_n^{c,\pm} = \mp\hbar\tilde{\omega} + \left\{\Delta^2 + (n+\frac{1\pm1}{2})2e\hbar B(v_c^2 + \Delta/\tilde{m})\right\}^{1/2}$$

(S3)

$$E_n^{v,\pm} = \pm\hbar\tilde{\omega} - \left\{\Delta^2 + (n+\frac{1\pm1}{2})2e\hbar B(v_c^2 + \Delta/\tilde{m})\right\}^{1/2}$$

Finally, we re-index the levels such that for $[n, \sigma = +1/2]$ becomes $[N = n+1, \sigma = +1/2]$ and $[n, \sigma = -1/2]$ becomes $[N = n, \sigma = -1/2]$, (see Suppl. Fig. 1), to finally get,

$$E_{N>0}^{c,\pm} = \mp\hbar\tilde{\omega} + \{\Delta^2 + 2eN\hbar B(v_c^2 + \Delta/\tilde{m})\}^{1/2}$$
$$E_0^c = \hbar\tilde{\omega} + \Delta$$

(S4)

$$E_{N>0}^{v,\pm} = \pm\hbar\tilde{\omega} - \{\Delta^2 + 2eN\hbar B(v_c^2 + \Delta/\tilde{m})\}^{1/2}$$
$$E_0^v = -\hbar\tilde{\omega} - \Delta$$

Note that in this indexing convention, ±refers to the effective spin, and the N=0 level is non-degenerate. This is expected for massive Dirac fermions. (see Suppl. Fig. 1)

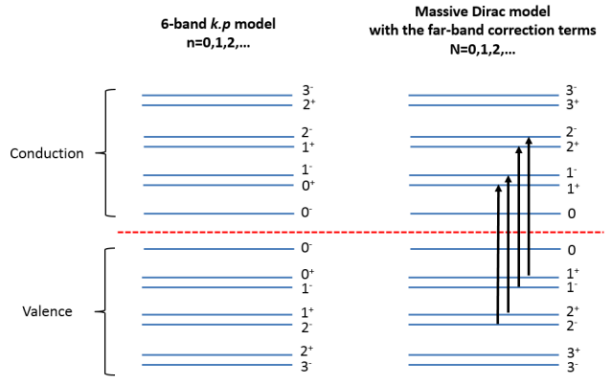

**Supplementary figure 1.** Landau level index in the 6-band **k.p** model and the massive Dirac model with far-band parameters (shown for the trivial case, for a deeper discussion of the effective Zeeman splitting in IV-VI semiconductor consult ref. 1 and 5).

The interband selection rules restrict us to transitions obeying $\Delta N = \pm 1 \; and \; \Delta \sigma = \pm 1$. Thus the interband transition energies all obey a quasi-massive Dirac model given by:

$$E_N^{c,\pm} - E_{N\pm1}^{v,\mp} = \sqrt{\Delta^2 + 2v_D^2 \hbar eBN} + \sqrt{\Delta^2 + 2v_D^2 \hbar eB(N \pm 1)} \qquad (S5)$$

Here we use the definition of the Dirac velocity $v_D^2 = v_c^2 + \Delta/\widetilde{m}$

The intraband transitions obey $\Delta N = \pm 1 \; and \; \Delta \sigma = 0$. The transition energy for a conduction band CR in the case is given by:

$$E_{N+1}^{c,\pm} - E_N^{c,\pm} = \sqrt{\Delta^2 + 2v_D^2 \hbar eB(N+1)} - \sqrt{\Delta^2 + 2v_D^2 \hbar eBN} \qquad (S6)$$

## S2. RHEED and AFM characterization

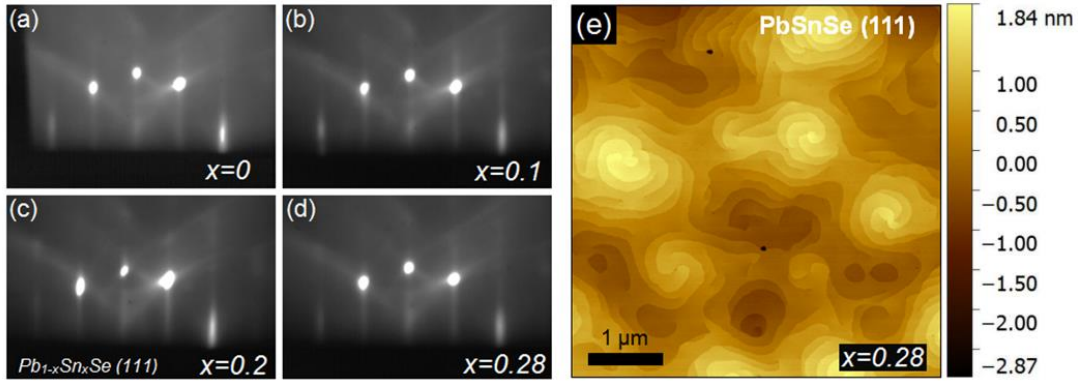

**Supplementary Figure 2.** (a-d) Reflection high-energy electron diffraction of $Pb_{1-x}Sn_xSe$ epilayers on $BaF_2$ (111) substrates recorded during MBE growth for *x* = 0, 0.10, 0.20 and 0.28 respectively. (e) An atomic force microscopy image of the *x* = 0.28 epilayer with 1 μm layer thickness.

Reflection high energy electron diffraction (RHEED) patterns recorded *in situ* during MBE growth of $Pb_{1-x}Sn_xSe$ films with various compositions are shown in Suppl. Fig. 2(a-d). For all layers, smooth 2D growth occurs after few nanometer deposition on $BaF_2$ (111), independently of the film composition. The high quality of the layer is evidenced by sharp diffraction spots on the Laue semicircle and intense Kikuchi lines arising from diffraction from bulk lattice planes. Similar RHEED patterns are observed for the $Pb_{1-x}Sn_xTe$ epilayers. No surface reconstruction was observed during deposition. The surface of the films is atomically flat, exhibiting only single monolayer steps of 3.52 Å height. This is exemplified by the atomic force microscopy (AFM) image presented in Suppl. Fig. 2(e). At the given growth temperature, growth proceeds in a 2D step-flow mode. Due to pinning of surface step at screw type threading dislocations originating from the $Pb_{1-x}Sn_x(Se,Te)$ / $BaF_2$ (111) lattice-mismatch of $\Delta a/a$ ~ 1.6% , a characteristic spiral step structure is formed.[6]

## S3. Magnetooptical data and analysis for $Pb_{1-x}Sn_xTe$

In order to consolidate the general aspect of our results we have also measured a number of Pb$_{1-x}$Sn$_x$Te samples and extracted the velocity and band gap using the same analysis that is shown for Pb$_{1-x}$Sn$_x$Se in the manuscript as well as in our previous work on Pb$_{0.54}$Sn$_{0.46}$Te.[7] Suppl. Fig. 3(a,b) show the magnetooptical Landau-level transitions extracted from the spectra shown in Suppl. Fig.3(c,d) for two respective samples having x=0.14 and x=0.56. A massive Dirac model is again used to extract the velocity (Suppl. Fig. 3(e)) and band gap (Suppl. Fig. 3(f)). Using $v_c^\ell = 7.4 \times 10^5 \, m/s$ for the longitudinal valley, and $v_c^o = 5.1 \times 10^5 \, m/s$ for the oblique valleys, we extract $\Delta/\tilde{m}$ and plot it versus x in Suppl. Fig. 3(g). $\Delta/\tilde{m} < 0$ points to a non-zero topological index above x≈0.40. This shows that the model is also valid for Pb$_{1-x}$Sn$_x$Te.

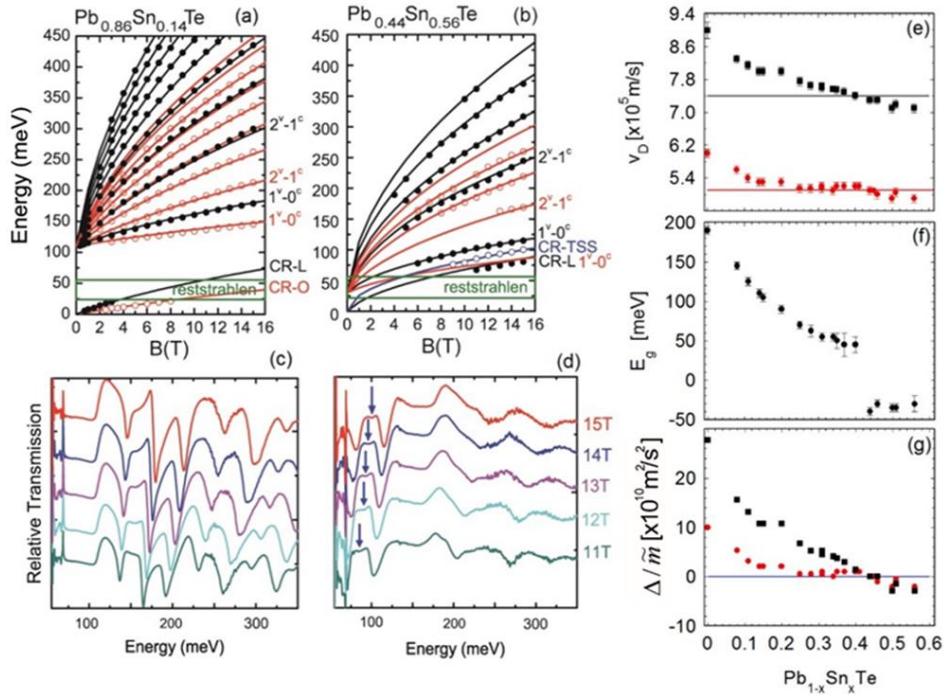

**Supplementary Figure 3**. (a,b) Magnetooptical Landau dispersions for Pb$_{1-x}$Sn$_x$Te x=0.14, and x=0.56 at T=4.5K. Circles denote data points extracted from mid- and far-IR magnetooptical spectra partially shown for each respective sample. Solid lines are curve fits using the massive Dirac model for bulk states and a massless Dirac model for the cyclotron resonance of the TSS (CR-TSS). Black is used for the longitudinal bulk valley, red is used for the oblique bulk valleys, and blue is used to denote the CR-TSS of the $\overline{\Gamma}$-Dirac cone. (c,d) Mid-IR magnetooptical transmission spectra measured between 11T and 15T in x=0.14 and x=0.56. A blue arrow points to CR-TSS of the $\overline{\Gamma}$-Dirac cone observed for x=0.56. Longitudinal (black) and oblique (red) velocity (e) and band gap (f) versus x. (g) $\Delta/\tilde{m}$ versus x.

It is worthwhile to note that the behavior of the band gap versus x in Pb$_{1-x}$Sn$_x$Te does not seem to be continuous. This can be interpreted as evidence for a first order phase transition.[8,9] It also could be due to short range alloy disorder as hinted in a previous ARPES study [8] but can also be the result of Bi-doping of Pb$_{1-x}$Sn$_x$Te when x>0.25. This is required to compensate the high native p-type hole concentration of Pb$_{1-x}$Sn$_x$Te that results from (Pb,Sn) vacancies for increasingly large Sn contents.[10]

**S4. The massive Dirac approximation and the value of $\tilde{m}$**

The values measured for $\tilde{m}/m_0$ are listed in table I in the text. Our values are in reasonable agreement with previous literature on PbSe and Pb$_{1-x}$Sn$_x$Se. It is important to check the validity of our assumptions that k$^4$ term in Hamiltonian (or equivalently those that vary as B$^2$ under the square root in the Landau levels) can be neglected. If we do not neglect any terms, the Landau levels are given by:

$$E_N^{c,\pm} = \mp\hbar\tilde{\omega} + \left\{\left(\Delta + \frac{N\hbar eB}{\tilde{m}}\right)^2 + 2eN\hbar B v_c^2\right\}^{1/2}$$

(S7)

$$E_N^{v,\pm} = \pm\hbar\tilde{\omega} - \left\{\left(\Delta + \frac{N\hbar eB}{\tilde{m}}\right)^2 + 2eN\hbar B v_c^2\right\}^{1/2}$$

For a typical value of $\tilde{m}/m_0 \approx 0.2$, we can safely neglect the B$^2$ term since it only becomes relevant for very-high indices (N>7) at very high fields. Our quantitative values are constrained mostly by the curve fits of low index Landau levels and are thus not affected by the B$^2$ term. We further confirm thus by comparing, in suppl. Fig. 4, fits using the full model to those using the massive Dirac model shown in the manuscript.

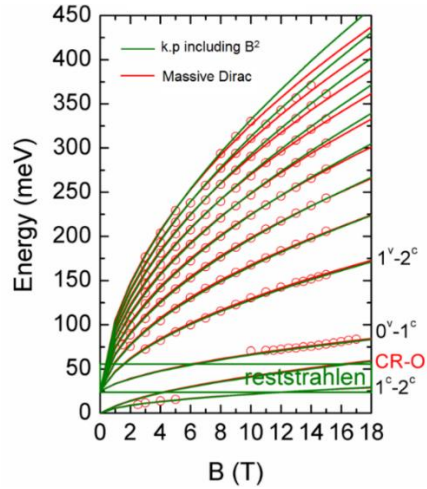

**Supplementary figure 4.** Magnetooptical transitions from the full **k.p** model that includes the B$^2$ term compared to those obtained using the massive Dirac model shown in the manuscript. We only show data and analysis for the oblique valleys for x=0.19 (at 4.5K).

The parameters obtained from the full model agree with those shown in the manuscript within experimental uncertainty (2Δ=-25±5meV, v$_c$=(4.6±0.1)x10$^5$m/s, $\tilde{m}/m_0 \approx 0.24$).

In any case, the $\tilde{m}/m_0$ terms yields a small correction to the band edge mass as can be seen for x=0.19 and x=0.24:

|  | $|\Delta|/v_c^2$ | $\tilde{m}$ | $m = \left|(-v_c^2/|\Delta| + 1/\tilde{m})^{-1}\right|$ |
|---|---|---|---|
| x=0.19 | 0.01m$_0$ | 0.24m$_0$ | 0.0104m$_0$ |
| x=0.24 | 0.02m$_0$ | 0.18m$_0$ | 0.022m$_0$ |

Supplementary table 1. Far-band corrections to the Dirac mass for x=0.19 and x=0.24. The impact of $\tilde{m}$ remains small.

Finally, in the case where B² terms cannot be neglected, our proposal to measure the topological index via the Dirac velocity in Landau level spectroscopy still holds, and Eq. S7 can be expanded and written as:

$$E_N^{c,\pm} = \mp\hbar\widetilde{\omega} + \left[\Delta^2 + 2eN\hbar B\left(v_c^2 + \frac{\Delta}{\widetilde{m}}\right) + \left(\frac{N\hbar eB}{\widetilde{m}}\right)^2\right]^{1/2}$$

$$E_N^{v,\pm} = \pm\hbar\widetilde{\omega} - \left[\Delta^2 + 2eN\hbar B\left(v_c^2 + \frac{\Delta}{\widetilde{m}}\right) + \left(\frac{N\hbar eB}{\widetilde{m}}\right)^2\right]^{1/2}$$

(S8)

$v_D^2 = v_c^2 + \frac{\Delta}{\widetilde{m}}$ can still be defined as before, and varies as expected when Δ changes sign.

**S5. Proposal for Hg$_{1-x}$Cd$_x$Te**

For Hg$_{1-x}$Cd$_x$Te, a similar expression can be obtained for the velocity of the $\Gamma_8$ band through the band inversion:

$$v_D^2 = v_c^2 + \Delta\left(\frac{\gamma_1 + 2\gamma}{m_0}\right)$$

For the Hg$_{1-x}$Cd$_x$Te $\Gamma_6$ band we find:

$$v_D^2 = v_c^2 + \Delta\left(\frac{1}{m_0} + \frac{v_c^2}{2\Delta + SO}\right)$$

where $2\Delta + SO$ is the $\Gamma_6$-$\Gamma_7$ energy separation. 2Δ is the energy separation between the $\Gamma_6$ and $\Gamma_8$ bands and γ and γ1 are far-band parameters as defined in ref. [11]. The critical velocity is defined as follows:

$$v_c = \sqrt{\frac{E_p}{3m_0}} \approx 1.06\times 10^6\, m/s$$

E$_p$ ≈ 19eV is the square Kane matrix element, for x=x$_c$=0.16 at 4.5K.[11,12]

The Dirac velocities are not the same for both bands, as a result of the band asymmetry of Hg$_{1-x}$Cd$_x$Te induced by the spin-orbit interaction. However, in the topological regime, when the gap is negative, both velocities are reduced.

Finally, it is important to note that the Landé g-factor is large in Hg$_{1-x}$Cd$_x$Te, reaching more than g=300 near the topological phase transition. This is expected for massive Dirac (and Kane) fermions where $g$ is on the order of $\frac{m_0}{m}$ the inverse band-edge mass. (See supplementary section S1 and suppl fig. 1 for a through discussion of the g-factor).